\documentclass{article}

  \usepackage[preprint]{neurips_2025}

\usepackage[utf8]{inputenc} 
\usepackage[T1]{fontenc}    
\usepackage{hyperref}       
\usepackage{url}            
\usepackage{booktabs}       
\usepackage{amsfonts}       
\usepackage{nicefrac}       
\usepackage{microtype}      
\usepackage{xcolor}         
\usepackage{multirow}       
\usepackage{array}          
\usepackage{amsmath}        %
\usepackage{graphicx}       %
\setcitestyle{authoryear, square, citesep={;}}  
\title{Game-Theoretic Modeling of Heterogeneous Investor Interactions for Stock Price Forecasting}

\author{%
  Yong Zhang\textsuperscript{1} \quad  Xinxiao Wu\textsuperscript{1} \quad Yunde Jia\textsuperscript{2} \quad  Che Sun\textsuperscript{2}\dag \\
  \textsuperscript{1}School of Computer Science,\ Beijing Institute of Technology\\
  \textsuperscript{2}Faculty of Engineering, Shenzhen MSU-BIT University\\
  \texttt{\{zhangy01, wuxinxiao\}@bit.edu.cn} \\
  \texttt{\{jiayunde,sunche\}@smbu.edu.cn}\\
}

\begin{document}

\maketitle

\begin{abstract}

Accurate stock price forecasting has consistently remained a pivotal yet challenging FinTech task that underpins quantitative trading and investment decision making. Recent efforts have been dedicated to modeling various complex relationships among stocks in the stock market toward more reliable stock price forecasting. These methods depend heavily on strong \textit{static prior assumptions} by modeling either temporal dependencies within individual stocks or spatial dependencies across different stocks based on predefined structures, while the complex market dynamics that drive stock price movements remain unexplored. To alleviate this issue, we propose a novel game-theoretic modeling method that captures heterogeneous investor interactions for stock price forecasting. The core idea is to embed game-theoretic mechanisms into the heterogeneous graph structure to finely model the dynamic strategic interactions among heterogeneous investors with respect to target stocks. Additionally, temporal positional encoding is adopted to reflect the differentiated influences of each game event at different time steps within the time window on future stock price movements. Leveraging heterogeneous graph networks, we proxy the intricate dynamics of the stock market through investor games and enable real-time information propagation and node updates among all nodes. Extensive experiments conducted on two real-world benchmark datasets demonstrate that our method effectively outperforms state-of-the-art stock price forecasting methods.

\end{abstract}

\section{Introduction}

In the stock market, stock selection and stock timing are the most concerned decision-making focuses of investors, all of which rely on the accurate prediction of the future stock prices of individual stocks, by which investors can maximize investment returns. The emergence of data science methods has provided researchers with new insights into stock price prediction. Methods such as autoregressive integrated moving average (ARIMA) \citep{ariyo2014stock}, principal component analysis (PCA)\citep{kelly2019characteristics}, random forest \citep{khaidem2016predicting}, and kalman filter \citep{yan2015application}, as well as their applications and integrations, have demonstrated remarkable success. These methods typically model the stock price prediction task as a time-series regression paradigm.

Early deep learning methods commonly treat stock-related volume-price indicator sequences as ordinary time series and adopt conventional sequential deep learning models \citep{chen2019investment} to model internal dynamics within stocks and predict stock prices, as exemplified by RNN \citep{li2019multi}, LSTM \citep{wang2019alphastock}, and Transformer \citep{xu2021relation}. Although such time series modeling methods have achieved promising progress in stock forecasting, they suffer from limited generalization ability on out-of-sample data. Modern approaches, in contrast, combine stock correlations beyond volume-price sequences to capture cross-stock influences \citep{cui2023temporal,qian2024mdgnn}.The relationships between stocks are inherently graph-like, making graphs the most intuitive and scientifically appropriate choice as an information carrier. The most typical relation graphs are pre-defined static graphs \citep{qian2024mdgnn}, data-driven correlation graphs \citep{yin2021forecasting} and temporally-aware hypergraphs \citep{feng2019temporal}.These methods have substantially boosted stock prediction performance by exploiting graph structures to model cross-stock interdependencies.

Despite these advances, two critical issues remain underexplored. First, the stock market is a complex chaotic system, where the driving forces of stock prices often stem not only from other related individual stocks but also from close interactions with diverse entities in the market. However, existing methods tend to confine themselves to modeling stock-to-stock relationships while neglecting the simultaneous modeling of multi-entity relationships, rendering them incapable of explaining extreme price fluctuations of individual stocks. Second, the structure of the stock market evolves dynamically over time, and predefined domain knowledge can hardly cover this dynamic evolution process, preventing relational modeling from achieving true real-time alignment with stock market dynamics.


In this paper, we propose a novel stock price forecasting model that effectively addresses both issues. Our method demonstrates  effectively enhancements by integrating
temporal dynamics learning and heterogeneous graph modeling in an end-to-end manner.
We employ a heterogeneous graph structure to fully model the multi-entity relationships in the stock market that drive price fluctuations, going beyond mere stock-to-stock correlations. Furthermore, we innovatively introduce game theory to conduct fine-grained modeling of capital flows and strategic interactions among market participants, so as to faithfully reflect the inherent dynamics of the stock market. Finally, the stock node representations updated via graph convolution are adopted for the final stock price prediction task. The main contributions of this paper are summarized as follows:

\begin{itemize}
   \item[$\bullet$]We propose a novel multi-stock price forecasting method based on game-theoretic mechanism termed GameStock, which could dynamically model the strategic and capital games among heterogeneous investors participating in stock market transactions in real time, and it is precisely these long-short games that are the fundamental driving forces behind stock price changes.
   \item[$\bullet$] Unlike conventional game-theoretic methods, GameStock fuses game-theoretic mechanism design with a heterogeneous graph network. By utilizing graph structures to model the intricate relationships among market entities, the model effectively channels gaming capital flows across individual stocks. Furthermore, the adoption of heterogeneous graph structures substantially strengthens the modeling of stock correlations in the market, and richer contextual information guarantees more robust and reliable relational learning.
   \item[$\bullet$] We introduce a sparse game-theoretic dataset designed to model strategic interactions among heterogeneous investors, underscoring the potential of this research direction for stock price forecasting. To facilitate further academic exploration, we will subsequently release this dataset publicly.
\end{itemize}

\section{Related work}
\label{gen_inst}

\subsection{Stock prices prediction}
In the field of financial time series analysis, stock price prediction has consistently been a focal point of research. Traditional methods primarily rely on statistical and econometric models, such as Autoregressive Integrated Moving Average (ARIMA) \citep{adebiyi2014comparison,wang2012stock} and the Vector Autoregressive (VAR) Model \citep{zivot2006vector}, focusing on numerical features \citep{piccolo1990distance,wang1996stock,tseng2002combining} and performing technical analysis based on volume-price indicator data. However, these methods exhibit limitations in feature extraction and poor performance in dealing with nonlinear and non-stationary time series data \citep{cui2023temporal}. The rapid advancement of deep learning has greatly facilitated stock price prediction, parts of the works following previous paradigm employ recurrent neural networks \citep{nelson2017stock,qin2017dual} or convolutional neural networks \citep{jiang2023re} to model a single stock price and predict its short-term trend. To capture richer and more fine-grained feature patterns to enhance prediction performance, some works have explored other techniques, such as the adversarial training \citep{feng2018enhancing}, self-attention mechanism \citep{li2018stock}, gated causal convolutions \citep{wang2021hierarchical}, and neural networks pre-training \citep{wang2025pre}. Meanwhile, to mitigate the adverse impact of noise interference from single data sources on stock price forecasting, some efforts have turned to multimodal learning. In particular, financial text data has been extensively incorporated into stock price prediction modeling, thereby elucidating intricate market insights that extend well beyond mere considerations of price dynamics, trading volumes, or financial indicators \citep{schumaker2009textual,si2013exploiting,du2024explainable,li2024causalstock,dong2024fnspid}. Given the inherently complex nature of stock sequences and their intrinsic multi-pattern characteristics, a number of studies have attempted to extract more sophisticated and diversified feature patterns. Techniques such as multi-scale temporal modeling \citep{liu2021multi,song2025multi,li2025cograsp}, multi-level learning \citep{zhang2025multi}, and mixture-of-experts systems \citep{liu2025mera} have been employed to capture the richest possible temporal information for improving stock price prediction.

\subsection{Graph Construction in Stock Prediction}
In addition to studies on individual stock dynamics, effectively modeling the correlation relationships among different stocks is regarded as another vital way to enhance stock price forecasting performance. The most prevalent strategy is to construct graph structures to characterize such inter-stock correlations, leverage graph neural networks (GNNs) to embed relational information, and subsequently feed these learned embeddings into downstream prediction modules for stock trend forecasting \citep{wang2021hierarchical}. Specifically, the relation graphs could be constructed based on existing concepts or predefined relationships such as stockholders \citep{chen2018incorporating}, industry-sector \citep{feng2019temporal,kim2019hats,sawhney2020spatiotemporal}, and concepts \citep{xu2021hist}. 
Nevertheless, predefined relationships and static graph structures severely constrain the effectiveness of relational modeling and fail to  align with the highly volatile nature of the stock market. Accordingly, a number of studies have shifted toward diversified stock relational modeling to alleviate this limitation, such as hidden conceptual relationships \citep{xu2021relation}, potential correlations between stocks and market sectors \citep{hsu2021fingat}, similarity-based dynamic latent relationships \citep{wang2022adaptive}, and implicit relevance hypergraph relations \citep{huynh2023efficient}, etc.
Despite the considerable progress achieved in previous works, the underlying market dynamics that fundamentally drive stock price fluctuations remain insufficiently explored.


\subsection{Game Theory-based Methods}
Game theory, as a classic branch of applied mathematics, aims to investigate how individuals interact with one another in competitive or cooperative scenarios, and has been widely applied in economics, biology, computer science and many other disciplines \citep{baltz2004behavioral,camerer2004advances,crawford1995theory}. Stylized games such as the Prisoner’s Dilemma, Rock-Paper-Scissors, Stag Hunt Game, and Ultimatum Game have been extensively analyzed in the context of multi-agent systems \citep{sreedhar2024simulating,akata2025playing,fan2024can,lore2023strategic}. Although theoretical research based on game theory has become fairly mature, the integration of game theory and deep learning for stock price forecasting still remains largely unexplored. In this paper, we innovatively adopt game-theoretic principles to model the strategic interactions among multiple agents in the stock market, so as to characterize the underlying market dynamics that fundamentally drive stock price fluctuations, and further enhance stock price forecasting performance via real-time information gain.

\section{Preliminary \& problem formulation}
\label{headings}

\subsection{Preliminary}


In our framework, we integrate game-theoretic strategies and heterogeneous graph networks for stock prices prediction. In this section, we introduce the fundamental concepts of game theory and the notations for heterogeneous graphs.
\paragraph{Heterogeneous Graph} Considering the complex entity relationships in the stock market, which go beyond the interactions among individual stocks, we adopt a heterogeneous graph architecture to model such intricate relations. Specifically, we incorporate three distinct types of entities as nodes to construct the heterogeneous graph, namely stocks, industries, and investors, which can be denoted as $\mathcal{G} = (\mathcal{V}, \mathcal{E}, \mathcal{A}, \mathcal{R})$, where each node $v \in \mathcal{V}$ is associated with a node type $\phi(v) \in \mathcal{A}$, and each heterogeneous edge $e \in \mathcal{E}$ corresponds to a relation type $\psi(e) \in \mathcal{R}$.

\paragraph{Game Theory} In our framework, all investors in the stock market are categorized into three types, namely institutions, hot money, and retail investors. The strategic and capital games surrounding the target stock are conducted among these three types of investors, which are regarded as players. The set of strategic actions available to them is denoted as $\mathit{Action} = \{\mathit{Buy}, \mathit{Sell}, \mathit{Hold} \ or \ \mathit{Bear\ position}\}$,

\subsection{Problem formulation}
Following the setup of existing works [Xia et al., 2024; Fan and Shen, 2024; Huynh et al., 2023], we formulate the stock price prediction problem into a stock relative price change prediction problem.

Given set of stocks $\mathit{S} = \{\mathit{s_1}, \mathit{s_2}, \cdots,  \mathit{s_N}\}$, each $\mathit{s_i}\subseteq \mathit{S} $ has historical trading data on trading day $\mathit{t}$ represented as the vector $\mathit{X_{S_i}^t}$.
 Additionally, Given a graph $\mathcal{G} = (\mathcal{V}, \mathcal{E}, \mathcal{A}, \mathcal{R})$ , where $\mathcal{V}$ denotes the set of entity vertices of all distinct types in the graph, and $\mathcal{E}$ represents the interactive relationships among different entities. We set a looking back windows $\mathit{L}$, our task in to use the relation graph $\mathcal{G}$ and the historical data of $\mathit{S}$ during day $\mathit{T}-(\mathit{L}-1)$ to 
$\mathit{T}$, donated as $\mathit{X}_{S}^{T-(L-1):T} = \{ \mathit{X}_{s_i}^{T-k}|i=1,2,\dots,N;k=L-1,L-2,\dots,0 \}$ , to predict the relative price change of all stocks in the stock pool $\mathit{S}$ on the next trading day $\mathit{T}+1$, donated as $\mathit{y_S^{T+1}}$. Mathematically, this can be formalized as
\begin{equation}\label{...}
\mathit{\hat{y}_S^{T+1}} = f(\mathit{X}_{S}^{T-(L-1):T},\mathcal{G};\Theta),
\end{equation}
where $\mathit{f}$ is our prediction approach and $\Theta$ represents the model parameters.






\begin{figure}[t]
    \centering
    \includegraphics[
        width=0.9\textwidth,
        trim=0 80 0 0,
        clip
    ]{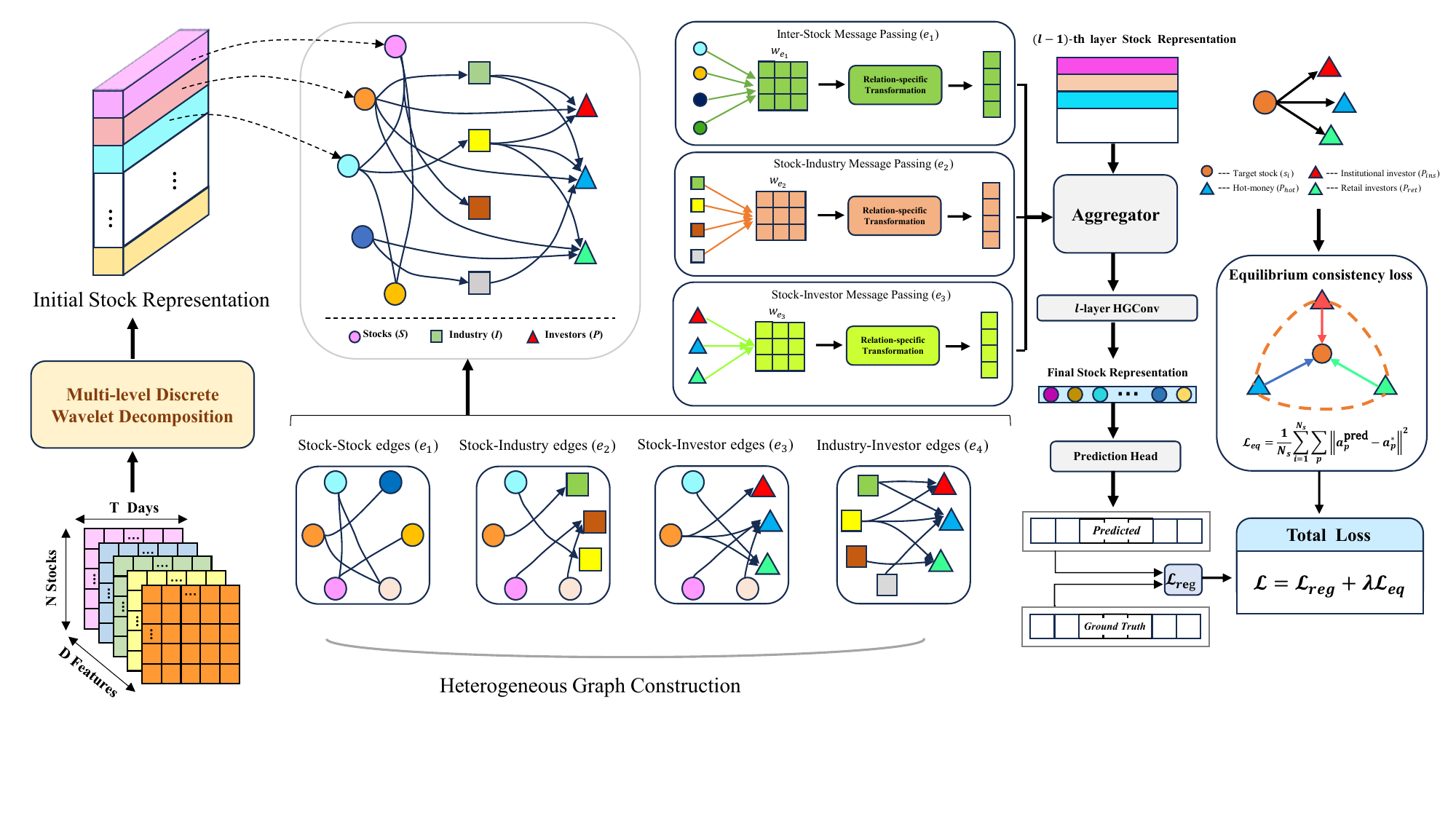}
    \caption{Overview of our GameStock architecture.}
    \label{fig:model}
\end{figure}

\section{Methodology}
\label{others}
As illustrated in Figure 1, our model is primarily composed of three key components: the multi-level discrete wavelet transform Module (M-DWT) , the heterogeneous graph convolutional network Module (HGCN) and the game-theoretic relation enhancement Module (GRE).

Our model initially applies M-DWT to identify periodic temporal trends. This method allows for the detection of inherent daily, weekly, and monthly patterns within stock sequences data. By utilizing the inherent downsampling property of wavelet filters, it can recognize temporal patterns across multiple frequency scales. HGCN is subsequently employed to model the diverse and complex entity relationships in the stock market,  whereas GRE dynamically augments these edge relations based on game-theoretic principles. The stock node representations, updated through the graph network, are then used for final stock price prediction.

\subsection{Multi-level Discrete Wavelet Transform Module (M-DWT)}

As demonstrated in Figure 1, this part is dedicated to recognizing
the periodic temporal patterns in stock indicator sequences, encompassing the intrinsic daily, weekly, and monthly patterns in stock prices. These patterns are respectively regarded as representing short-term speculation, medium-term swing movements, and long-term trends that drive stock price fluctuations. For each stock $s_i$ , the historical daily indicator data over the preceding 
$L$ time steps (i.e., from  $\mathit{T}-(\mathit{L}-1)$ to $\mathit{T}$) is donated as $\mathit{X}_{s_i}^{T-(L-1):T}$ $\in$ $\mathbb{R}^{L \times D}$, 
where each indicator sequence is treated as an independent feature channel. After standardizing these indicator sequences separately, the whole set is fed into the DWT module to obtain the coefficients of different frequency temporal patterns for each indicator sequence as follows:
\begin{equation}\label{...}
\{cA_{j,l}, cD_{j,k}\}= DWT( \mathit{X}_{s_i}^{T-(L-1):T},l), \:\: k \in [1,2,\dots,l ]
\end{equation}
where, $l$ denotes the decomposition level of the discrete wavelet transform, $cA_{j,l}$ represents the final approximation coefficient sequence of the $j$-th indicator sequence after $l$-level wavelet decomposition, and $cD_{j,k}$(with 
$k$=1,2,…,$l$) represents the $l$ detail coefficient sequences obtained from the same decomposition. In the subsequent experiments of this paper, we employ the standard orthogonal Daubechies (db) wavelet basis to perform the wavelet decomposition and treat the decomposition level $l$ as an adjustable model parameter. 

\paragraph{Temporal Attention} Studies have shown that feature patterns at different frequencies exert distinct impacts on future stock prices. To this end, we employ a temporal attention mechanism $\zeta(\cdot)$ which learns to weigh critical frequency pattern that impact future prices. This mechanism aggregates temporal hidden states $h_k$ from different frequency ranges into an overall representation using learnt attention weights $\alpha_k$ for each
decomposition level $k$ . We formulate this mechanism as:
\begin{equation}\label{...}
h_k= Linear(||_{j=1}^{D}MaxPool(cD_{j,k}))
\end{equation}
\begin{equation}\label{...}
\zeta(H)=\sum\limits_k \alpha_k h_k,\; \alpha_k=\frac{exp(e_k)} {\sum\limits_{k}exp(e_k)}, \; e_k=W_2 \cdot\sigma(W_1 h_k+b_1)+b_2
\end{equation}

where, $W_1$ and $W_2$ are learnable linear transform matrices, $b_1$ and $b_2$ are bias terms. $\sigma$ denotes a non-linear activation function. $\alpha_k$ represents the learnt attention weights used to aggregate all temporal features across different levels while assigning higher weights to important hierarchical patterns.

\paragraph{Trend-Fluctuation Fusion} In the stock market, stock price movements are mainly governed by both low-frequency long-term trends and high-frequency short-term fluctuations. Economic environment shifts, sector-specific policy incentives, and unforeseen news events all significantly influence future stock prices, yet their impacts differ across individual stocks. To address this, we propose a trend-fluctuation fusion mechanism that adaptively integrates the long-term trends and short-term fluctuations of different stocks, thereby generating the final representation for each individual stock.
\begin{equation}\label{...}
h_{trend}= Linear(||_{j=1}^{D}AvgPool(cA_{j,l}))
\end{equation}
\begin{equation}\label{...}
\lambda=Sigmoid(Linear(h_{trend})
\end{equation}
\begin{equation}\label{...}
z_{s_i}= h_{trend} + \lambda \cdot \zeta(H)
\end{equation}
where $\lambda$ denotes a learnable parameter used to adaptively balance the influence of short-term fluctuations. We concatenate representations of all stocks to form $\boldsymbol{Z}\in \mathbb{R}^{N \times M}$.

\subsection{Heterogeneous graph convolutional network Module (HGCN)}

\paragraph{Heterogeneous graph Construction} We capture the interplay among stocks, sectors, and investors via a heterogeneous graph, modeling three entity types within the stock market. Specifically, we denote directed and
labeled multi-graphs as $\mathcal{G} = (\mathcal{V}, \mathcal{E}, \mathcal{A}, \mathcal{R})$ with nodes (entities) $v \in \mathcal{V}$ and labeled edges (relations) $e \in \mathcal{E}$, where $\psi(e) \in \mathcal{R}$ is a specific relationship type. For convenience, we denote the three different types of nodes as stock node set $N_s=\{s_1,s_2,\dots,s_{|N_s|}\}$ , industry node set $N_i=\{i_1,i_2,\dots,i_{|N_i|}\}$, and investor node set $N_p=\{p_1,p_2,\dots,p_{|N_p|}\}$, respectively. Furthermore, we inject domain knowledge by constructing heterogeneous edge between entities based on two types of relations: industry and investor holding relations.

 \textit{Industry heterogeneous edges}: Stocks belonging to same industries,
collectively experience similar price trends based on the industry’s performance \citep{livingston1977industry}. To leverage this signal, we define relations between stocks as per the CSIINDEX standard. Formally, we construct heterogeneous edges
$e$ that connects stocks that belong to the same industry. In addition, heterogeneous edges are similarly established to connect stock nodes with their corresponding sector nodes, thereby modeling the relationships between stocks and their respective industries.

\textit{Investor holding heterogeneous edges}: Distinct holding preferences are observed among different investor types. Specifically, institutional investors exhibit a preference for blue-chip and high-quality growth stocks, whereas speculative hot money tends to concentrate on short-term speculative opportunities and technology-focused themes. Meanwhile, retail investors often follow the momentum of speculative capital, directing their attention to the same short-term trends and market narratives. Such heterogeneity in holding preferences serves as a critical factor driving the long-term divergence of stock price trends. For instance, large-cap bellwether stocks such as the Industrial and Commercial Bank of China (ICBC) and Kweichow Moutai, which are favored by institutional investors, are represented by heterogeneous edges.

We combine these relations to construct the heterogeneous graph $\mathcal{G}$ with all edges stored in a adjacency matrix $M$ $\in$ $\mathbb{R}^{|\mathcal{V} |\times|\mathcal{V}|}$, with entries $m(v_i,v_j)$ defined as 
\begin{equation}
\left.m(v_i,v_j)=\left\{
\begin{array}
{cc}1, & \quad (v_i, v_j)\in e \\
0, & \quad (v_i, v_j)\notin e
\end{array}\right.\right.
\end{equation}

Motivated by the architectures of R-GCN \citep{schlichtkrull2018modeling}, we employ the following simple propagation model for calculating the forward-pass update of an entity or node denoted by $v_i$ in a multi-relational heterogeneous graph:

\begin{equation}
h_{v_i}^{(l+1)}=\sigma\left(\sum_{e\in\mathcal{E}}\sum _{v_j \in \mathcal{N}_{v_i}^e}        \frac{1}  {c_{v_i,e}} W_r^{(l) }h_{v_j}^{(l)} + W_0^{(l)}h_{v_i}^{(l)} \right),
\end{equation}

where $\mathcal{N}_{v_i}^e$ denotes set of neighbor indices of node $v_i$ under relation $e\in\mathcal{E}$ . $c_{v_i,e}$ is problem-specific normalization constant that can either be learned or chosen in advance (such as  $c_{v_i,e}=|\mathcal{N}_{v_i}^e|$  ).

\subsection{Game-theoretic relation enhancement Module (GRE)}
As a critical source of information reflecting intraday capital flows, the post-market "Dragon and Tiger List" of A-shares captures the concentrated manifestation of capital gaming among different investors around individual stocks. The event of a stock being listed on the Dragon and Tiger List reflects short-term capital enthusiasm and speculation on that stock, which exerts either a positive or negative impact on its price over a subsequent period. However, this impact gradually diminishes over time until it eventually decays to zero. Motivated by this, we incorporate a game-theoretic mechanism to model the capital gaming among different investors around individual stocks, where investors act as players whose feasible strategies comprise buying, selling, holding,  or remaining in a short position. In the subsequent experiments, these strategies are encoded as -1, 1, and 0, representing selling, buying, and holding (or remaining in a short position), respectively.

For a given stock $s_i$, we formally define the set of Dragon and Tiger List events occurring within the window period as $E_{s_i}=\{(d_j, s_{d_j,s_i})\}$, where $d_j\in [t-W+1,t]$ denotes the specific date on which the Dragon and Tiger List event occurs for stock $s_i$, and $s_{d_j,s_i}$ represents the gaming triple associated with stock $s_i$ on that date, and $j$ counts the occurrences of stock $s_i$ on the Dragon and Tiger List. We attempt to encode each Dragon and Tiger List event to generate the corresponding game signal. Considering the considerable randomness of individual stocks appearing on the Top List, we introduce temporal positional encoding to reflect the differentiated impact of Top List events at different time steps on future predictions. The complete process can be expressed as:

\begin{equation}
\alpha_{(d_j,s_{d_j,s_i})}=\frac{\omega_{d_j}}{\sum_{(d_j,s_{d_j,s_i})\in E_{s_i}}\omega\prime} , \quad  \omega_{d_j}=e^{-\alpha\cdot(t-d_j)}
\end{equation}
\begin{equation}
v_{s_i}=\mathrm{MLP}\left([\mathrm{pos\_emb}(d_j);\mathbf{s}_{d_j,s_i}]\right)
\end{equation}
\begin{equation}
g_{s_i}=\sum_{(d_j,s_{d_j,s_i})\in E_{s_i}}\alpha_{(d_j,s_{d_j,s_i})} \cdot v_{s_i}
\end{equation}
\begin{equation}
H_s^{(L)}=z\odot h_s^{(L)}+(1-z)\odot g ,  \quad  z=\sigma(W_z\cdot[h_s^{(L)};g]+b_z)
\end{equation}

where $\omega_{d_j}$ denotes the exponentially decaying weight over time, characterizing the fact that earlier Dragon and Tiger List game events exert weaker impacts on future stock price prediction, which is consistent with our general intuition. Considering that an individual stock may appear on the Dragon and Tiger List multiple times within the window, we subsequently perform normalization on the weights. We employ temporal positional encoding to encode the listing dates, and integrate the information of game triples to generate the representation $v_{s_i}$ for each game event. Finally, based on the normalized weights above, we generate the game signal $g_{s_i}$ for each stock within the lookback window.

In our setting, gaming occurs exclusively among three types of heterogeneous investors i.e., the set of game players is defined as $N_{player}=\{p_{_{ins}},p_{_{hot}},p_{_{ret}}\}$, which respectively represent institutional investors, hot money (or speculative funds), and retail investors. We reasonably assume that investors of the same type exhibit convergent behavior in the stock market and therefore treat each player type as a cohesive whole for representation.  The strategies available to each player are drawn from the action set $\pi=\{-1,0,1\}$, corresponding to net selling, holding, and net buying behaviors, respectively. The equilibrium state is then given by : 
\begin{equation}
u_p(a_p,a_{-p},r)=a_p\cdot r+\lambda\cdot a_p\cdot\sum_{q\neq p}\beta_{p,q}\cdot a_q
\end{equation}
\begin{equation}
u_{p_{_{ins}}}(x^*,y^*,z^*,r)\geq u_{p_{_{ins}}}(x^{\prime},y^*,z^*,r),\quad\forall x^{\prime}\in\{-1,0,1\}
\end{equation}

\begin{equation}
u_{p_{_{hot}}}{(x^*,y^*,z^*,r)}\geq u_{p_{_{hot}}}{(x^*,y^{\prime},z^*,r)},\quad\forall y^{\prime}\in\{-1,0,1\}
\end{equation}

\begin{equation}
u_{p_{_{ret}}}{(x^*,y^*,z^*,r)}\geq u_{p_{_{ret}}}{(x^*,y^*,z^{\prime},r)},\quad\forall z^{\prime}\in\{-1,0,1\}
\end{equation}

\begin{equation}
\mathcal{L}_{\mathrm{eq}}=\frac{1}{N_s}\sum_{i=1}^{N_s}\sum_p\left\|a_p^{\mathrm{pred}}-a_p^*\right\|^2
\end{equation}

Here, $u_p(\cdot)$ denotes the payoff of each type of investor p when selecting a specific strategy $a_p$, r represents the 1-day return of the stock under gaming, $\beta_{p,q}$ is a coefficient measuring the degree of follow-up among different investors, and $(x^*,y^*,z^*)$ are the strategies chosen by the three types of investors in the equilibrium state.

\subsection{Loss Function}
We employ two heterogeneous graph convolutions (RGCNConv($\cdot$)) with an ELU activation between the first and second layer. Then we make the final prediction based on the updated node representations of stock nodes following the $L$-th  convolutional layers.
\begin{equation}
\hat{y}=\sigma(LayerNorm(H_{s}^{(L)})),
\end{equation}
We optimize our model using a combination of a pointwise regression and equilibrium consistency loss to minimize the difference between the predicted and actual return ratios while preserving a dynamic equilibrium state as:
\begin{equation}
\mathcal{L}_{\mathrm{pred}}=\frac{1}{N_s}\sum_{i=1}^{N_s}\left\|\hat{y}_i-y_i\right\|^2
\end{equation}
\begin{equation}
\mathcal{L}_{total}=\mathcal{L}_{pred}+\lambda\cdot \mathcal{L}_{eq}
\end{equation}

\section{Experiments}
In this section, we study our approach with comprehensive experiments, aiming to answer the research questions:

\begin{itemize}
   \item[$\bullet$] RQ1: How does our proposed approach perform compared with the state-of-the-art methods?
   \item[$\bullet$] RQ2: How is the effect of different modules in our approach?
   \item[$\bullet$] RQ3: Whether the game-theoretic mechanism module can capture more fine-grained inter-stock relationships compared with using graph structures alone, and thereby enhanced stock price prediction performance?
\end{itemize}

\subsection{Experimental setup}
\paragraph{Datasets:} We evaluate our approach based on three real world
datasets from the Chinese stock market. The statistics of the datasets are in Table 1. The CSI300 Index tracks the 300 largest and most liquid A-shares, reflecting overall market trends, while the CSI500 Index covers the next 500 largest stocks (excluding CSI300), representing small and mid-cap performance. Our dataset includes common volume-price indicators and technical indicators  for each individual stock, covering opening price, closing price, highest price, lowest price, trading volume (OHLCV), as well as 5-day, 10-day, 20-day, and 30-day moving average prices, all spanning from January 2017 to December 2024. We partition this dataset chronologically into three segments: a training period spanning from 2017 to 2022, a validation period spanning from 2022 to 2024,
and a testing period spanning from 2024 to 2024. Stocks included in the dataset were screened to ensure that those retained were present for at least 95\% of trading days over the entire sample period. The model that performs best on the validation set is used for final testing.

\begin{table}[htbp]
\centering
\caption{Statistics of datasets.}
\label{tab:dataset_stats}
\begin{tabular}{l c c c c}
\hline\hline 
Datasets & Stocks & Train Days       & Valid Days      & Test Days       \\
\midrule
CSI300   & 218    & 01/03-08/05 (1360) & 08/08-03/14 (389) & 03/15-12/31 (194) \\
CSI500   & 342   & 01/03-08/05 (1360) & 08/08-03/14 (389) & 03/15-12/31 (194) \\
CSI1000I   & 735  & 01/03-08/05 (1360) & 08/08-03/14 (389) & 03/15-12/31 (194) \\
\hline\hline 
\end{tabular}
\end{table}

\paragraph{Implementation Details:}Experiments are based on PyTorch and PyG. All models were implemented using PyTorch 1.12.1 and trained and evaluated on a single NVIDIA GeForce RTX 3090 GPU for 300 epochs, with an early stopping mechanism of 20 epochs introduced to prevent overfitting. The lookback window was set to 20, the learning rate was set to 0.001, and an L2 regularization of 0.001 was applied. During the training process, the ReduceLROnPlateau strategy was used to dynamically adjust the learning rate. In our proposed model, the standard orthogonal Daubechies wavelet is adopted as the wavelet basis, the number of wavelet decomposition levels is set to 3, and the embedding dimension is 48. The number of layers of the GAT was set to 2 layer and the number of hidden units was 64. In addition, each model was repeated 5 times to verify its stability, and the average performance was reported.

\paragraph{Baselines:} We compared the performance of our method with several stock price prediction models: (1) LSTM [Hochreiter and Schmidhuber, 1997]: Specialized recurrent neural network that can capture long-term dependencies. (2) DA-RNN [Qin et al., 2017]: Attention-based LSTM model that enhances predictive power by integrating attention mechanisms.
(3)Adv-ALSTM [Feng et al.,2019] Tries to improve ALSTM through adversarial training to capture the stochastic nature of stock price and ameliorate the over-fitting.
(4) FEDformer[Zhou et al.,2022] Transformer-based model that leverages frequency-domain attention and mixture of experts for seasonal-trend decomposition to better capture the global properties of time series.
(5) HIST [Xu et al., 2021] Focus on the relevance between stocks and concepts, while utilizing both the stock’s shared information and individual information to improve the stock forecasting performance. 
(6) SFM [Zhang et al.,2017]: Decomposes the hidden states of LSTM memory units
into multiple frequency components to simulate the various underlying trading patterns behind stock price fluctuations.
(7) MASTER [Li et al.,2024]: By alternately performing intra-stock and inter-stock information aggregation to model the momentary and cross-time stock correlations, and leveraging market information for automatic feature selection.

\begin{table}[t]
\centering
\caption{Main results of GTHGAT and baselines for stock price forecasting task on different datasets. The best results are in bold and the second-best results are underlined (p<0.01).}
\label{Table2}
\resizebox{\textwidth}{!} {
\begin{tabular}{cccccccccc}

\hline
\noalign{\vspace{3pt}}
\multicolumn{2}{c}{\multirow{2}{*}{Model}}& \multicolumn{4}{c}{CSI300}&\multicolumn{4}{c}{CSI500}\\

\cmidrule(lr{0.5em}){3-6}  
\cmidrule(l{0.5em}r){7-10} 
\noalign{\vspace{3pt}}
\multicolumn{2}{c}{}& IC  & RankIC  & ICIR & RankICIR & IC  & RankIC  & ICIR & RankICIR \\
\hline

\noalign{\vspace{3pt}}
\multirow{4}{*}{} & LSTM  & 0.0233 & 0.0224 &0.1346  & 0.0968 &	0.0274 & 0.0262 & 0.1350 & 0.1086 \\

\multirow{4}{*}{} & DA-RNN  & 0.0227 & 0.0236	& 0.1301	& 0.0844	& 0.0249   & 0.0230	& 0.1308	& 0.0869\\

\multirow{4}{*}{} & Adv-ALSTM & 0.0245	& 0.0257	& 0.1362	& 0.1088	& 0.0287	& 0.0275	& 0.1395	& 0.1103\\

\multirow{4}{*}{} & FEDformer &0.0310	&0.0232	&0.1456	&0.1234	&0.0325	&0.0255	&0.1482	&0.1246\\

\multirow{4}{*}{} & HIST &0.0283	&0.0227	&0.1369	&0.1106	&0.0314	&0.0268	&0.1397	&0.1135\\

\multirow{4}{*}{} & SFM &0.0295	&0.0197	&0.1588	&0.1276	&0.0306	&0.0199	&0.1579	&0.1268\\

\multirow{4}{*}{} & MASTER &0.0312	&0.0275	&0.1428	&0.1125	&0.0334	&0.0280	&0.1436	&0.1142\\


\multirow{4}{*}{} & Ours &0.0325	&0.0251	&0.1432	&0.1187	&0.0345	&0.0276	&0.1435	&0.1214\\

\hline
\end{tabular}
}  
\label{table_MAP}
\end{table}

\paragraph{Metrics:} Following [Li et al., 2024], we employ four IC and
its variants metrics which are widely accepted in stock price prediction research [Lin et al., 2021]: IC, RankIC, ICIR, and RankICIR. IC represents the Pearson correlation coefficient and is calculated daily to assess the linear relationship between predicted and actual values. RankIC uses the Spearman
rank correlation coefficient to evaluate the monotonic relationship between predictions and actual outcomes. ICIR normalizes the IC by dividing it by its standard deviation, providing a measure of the consistency and reliability of the
IC. RankICIR Enhances the RankIC by normalizing it against its standard deviation, thereby offering an assessment of the RankIC’s consistency and reliability over time.

\subsection{Performance Comparison}
In Table 2, we compare ours model with the baseline methods, where ours model  consistently surpasses the vast majority of baselines across all evaluation metrics. Most baselines were evaluated using their original configurations and the same loss function for fairness. We discover that StockMixer demonstrates its superiority over LSTM, ALSTM, and FEDformer by time mixing to better capture multi-scale temporal information, suggesting that leveraging different patterns information from historical stock data can significantly enhance stock price prediction capabilities. Furthermore, while HIST aim to improve predictive accuracy by analyzing inter-stock relations, they rely on pre-defined and fixed graph structures, which may not effectively represent the complex interrelations among stocks in the market, leading to their underperformance compared to our method. The exceptional effectiveness of our method primarily stems from two key factors: 1) Our method fully considers the fluctuation patterns at different frequencies in the stock price-volume time series and achieves their effective fusion, ensuring that the model can capture multi-scale information to enhance prediction performance. 2) Our method innovatively models the relationships among diverse and complex entities in the stock market, rather than merely the simple correlation between stocks, which ensures that our method can integrate richer multi-source information for future stock price prediction. 3) By introducing game theory, our method realizes the dynamic enhancement of relationship modeling, which greatly improves the effectiveness of relationship modeling.

\subsection{Ablation study}
To validate the effectiveness of each component, we conducted ablation experiments, we have the following three variants. 1) \textbf{w/o M-DWT}: A single LSTM is used to replace the M-DWT module for processing time series data, while keeping all other components unchanged; 2) \textbf{w/o GRE}: Removing the GRE module from the original model, with all other modules kept unchanged; 3) \textbf{w/o HGCN+GRE}: The prediction is made solely based on the M-DWT module.

We observed that as the number of components included in each variant increased, the performance of the variants progressively improved, providing evidence for the positive impact of each component.

\begin{table}[t]
\centering
\caption{Ablation study results on different datasets.}
\label{Table3}
\resizebox{\textwidth}{!} {
\begin{tabular}{cccccccccc}

\hline
\noalign{\vspace{3pt}}
\multicolumn{2}{c}{\multirow{2}{*}{Ablation Variants}}& \multicolumn{4}{c}{CSI300}&\multicolumn{4}{c}{CSI500}\\

\cmidrule(lr{0.5em}){3-6}  
\cmidrule(l{0.5em}r){7-10} 
\noalign{\vspace{3pt}}
\multicolumn{2}{c}{}& IC  & RankIC  & ICIR & RankICIR & IC  & RankIC  & ICIR & RankICIR \\
\hline

\noalign{\vspace{3pt}}
\multirow{4}{*}{} & LSTM  & 0.0233 & 0.0224 &0.1346  & 0.0968 &	0.0274 & 0.0262 & 0.1350 & 0.1086 \\

\multirow{4}{*}{} & w.o.M-DWT  &0.0306	&0.0237	&0.1411	&0.1165	&0.0320	&0.0274	&0.1426	&0.1195\\

\multirow{4}{*}{} & w.o.GRE &0.0287	&0.0229	&0.1385	&0.1126	&0.0307	&0.0269	&0.1408	&0.1169\\

\multirow{4}{*}{} & w.o.HGCN+GRE &0.0256	&0.0226	&0.1357	&0.1079	&0.0283	&0.0264	&0.1365	&0.1137\\

\multirow{4}{*}{} & Ours &0.0325	&0.0251	&0.1432	&0.1187	&0.0345	&0.0276	&0.1435	&0.1214\\

\hline
\end{tabular}
}  
\label{table_MAP}
\end{table}

\subsection{Results of explainability}

To assess the effectiveness of our graph construction, we performed network analysis and comparative experiments. Beyond conventional graph structures, we construct a heterogeneous graph by incorporating market participants, i.e., investors, into the graph construction scope. Specifically, we obtain the holding preferences of different investor types for stocks and sectors from the East Asia Wealth Data Center. Based on this, we construct a heterogeneous graph structure, and we compared our method against the widely used industry sector graph. Compared to the industry sector graph, the
heterogeneous graph contains significantly more information about stocks but is not as over-densely connected as a correlation graph, where every pair of nodes is linked. A single connected component indicates that stocks are in a giant interconnected structure, with more frequent and diverse relations between stocks. Additionally, the significantly higher closeness centrality further suggests that the heterogeneous graph more comprehensively captures the relations related to stocks within the market. In contrast, the industry sector graph’s dispersed structure and lower connectivity result in a relatively limited amount of contained information while the correlation graph is over-connected reducing the reliability of the influence between stocks.

\section{Conclusion}
In this paper, we proposed a novel model for accurate stock price forecasting by capturing market dynamics and complex entity relationships. The model incorporates a wavelet transform module with a hierarchical attention mechanism to precisely identify fluctuation patterns and trend patterns in stock volume-price time series and enhance stock temporal representations. Furthermore, by introducing game theory equilibrium and constructing a well-designed heterogeneous graph structure, the model effectively captures complex interactions among different entities in the real stock market, while fully accounting for capital competition and strategic interactions among various investors over individual stocks, thereby significantly strengthening heterogeneous relation modeling. Extensive experiments conducted on Chinese A-share market data demonstrate that our proposed model outperforms state-of-the-art methods in both prediction performance and portfolio stability. Additionally, ablation studies and detailed analyses validate the effectiveness of the proposed components. In future works, we will focus on improving the model's robustness, particularly during periods of market instability, to maintain consistent reliability under various market conditions.

\begin{ack}


This work was supported by research grants to be acknowledged in  the final version.
\end{ack}


\bibliographystyle{plainnat}
\bibliography{ref}


\medskip

\end{document}